\documentclass[%
       draft,
        notitlepage,
        narroweqnarray,
        inline,
        ]{ieee}
\usepackage{ieeefig}
\usepackage{amsmath,graphics,graphicx,epsf,epsfig, lipsum} 
\usepackage{amssymb,layout,bm} 
\usepackage{amssymb,boxedminipage} 

\newtheorem{theorem}{Theorem}[section]

\begin{document}
\pagestyle{empty}
%----------------------------------------------------------------------
% Title Information, Abstract and Keywords
%----------------------------------------------------------------------
%\title[Short Title]{%
%       Long Title}
%
\title[ ]{%
      \huge{Interference Alignment Under Limited Feedback for MIMO Interference Channels}}

%\author[SHORT NAMES]{
%           		 Rajesh T Krishnamachari, 
%    			 \authorinfo{
%     			 Rajesh T K is with the Department of Electrical Engineering, University of Colorado, Boulder C0, 80305, USA,
%    			email: \mbox{krishnrt@colorado.edu}
%					  }
% \and
%    			Mahesh K Varanasi 
%     			\authorinfo{
%      			Mahesh K Varanasi is with the Department of Electrical Engineering, University of Colorado, Boulder C0, 80305, USA,
%  			 email: \mbox{varanasi@colorado.edu}
% 					}
%				 }

\author{Rajesh T. Krishnamachari \thanks{This work is supported in part by NSF Grant  CCF-0728955. The authors are with the Department of Electrical, Computer and Energy Engineering, University of Colorado, Boulder, CO 80309-0425 USA
   (e-mail:\texttt{ krishnrt@colorado.edu; varanasi@colorado.edu}).} \quad Mahesh K. Varanasi }  
\confplacedate{UnknownCity, UnknownState, June 28 -- July 3, 2009}
\maketitle
\thispagestyle{empty}

 %-----------------------------------Start of Abstract ---------------------------%
\begin{abstract} 
While interference alignment schemes have been employed to realize the full multiplexing gain of $K$-user interference channels, the analyses performed so far have predominantly focused on the case when global channel knowledge is available at each node of the network. This paper considers the problem where each receiver knows its channels from all the transmitters and feeds back this information using a limited number of bits to all other terminals. In particular, channel quantization over the composite Grassmann manifold is proposed and analyzed. It is shown, for $K$-user multiple-input, multiple-output (MIMO) interference channels, that when the transmitters use an interference alignment strategy as if the quantized channel estimates obtained via this limited feedback are perfect, the full sum degrees of freedom of the interference channel can be achieved as long as the feedback bit rate scales sufficiently fast with the signal-to-noise ratio. Moreover, this is only one extreme point of a continuous tradeoff between achievable degrees of freedom region and user feedback rate scalings which are allowed to be non-identical. It is seen that a slower scaling of feedback rate for any one user leads to commensurately fewer degrees of freedom for that user alone. 
\end{abstract} 
 %-----------------------------------End of Abstract ---------------------------%

% Do the keywords
\begin{keywords}
Composite Grassmann manifold, finite-rate feedback, interference alignment, interference channel, MIMO, quantization.
\end{keywords}

% -----------------------------------Start of Introduction ---------------------------%
\section{Introduction}\label{Introduction} 
     
     The importance of the role played by interference management in wireless networks has prompted many researchers to analyze the interference channel from an information-theoretic perspective.  While the capacity region remains unknown, many insightful characterizations have been developed through outer bounds \cite{MotahariKhandani,ShangKramerChan} and approximations \cite{EtkinTseWang}. The conventional wisdom was that if the strength of the interference is comparable to the actual signal, orthogonalizing the signaling dimensions between users was best providing  $\frac{1}{K}$ degrees of freedom per user in a $K$-user interference channel. Abandoning this `cake-cutting' approach however, \cite{CadambeJafar} demonstrated the achievability of the sum degrees of freedom for this channel of $\frac{K}{2}$ in this interference regime through what has come to be known as interference alignment (IA).  This surprising result has spurred further research in this area, including in other aspects of the scheme as in \cite{GrokopTseYates} in analyzing limited multipath and variation with number of users in the system 
     as well as in its applications as in \cite{TreschGuilland} in mitigating inter-cell interference in a cellular network. In the specific setup of frequency-selective channels, \cite{ShenHostmadsenVidal} and \cite{ChoiJafarChung} have suggested efficient schemes improving upon the original IA scheme in terms of data rates and multiplexing gains achievable through coding over a finite number of channel realizations.  Finally, while the notion of degrees of freedom is an academic construct, experimental verification of the benefits accruing from the interference alignment scheme  is available from several papers, such as \cite{GollakotaPerliKatabi}, where a combination of interference alignment and cancellation is shown to provide a significant throughput increase. 
          
The requirement of perfect channel state information at the transmitters (CSIT) by the IA scheme in \cite{CadambeJafar} is, of course, practically unrealizable  for a time-variant or frequency-selective system and this issue has recently begun to receive considerable attention (\cite{TreschGuilland,MotahariGharanMaddahAliKhandani,ThukralBolcskei}).  In particular, \cite{TreschGuilland} analyzed the impact of imperfect channel knowledge of the sum mutual information achieved by interference alignment, when applied to the downlink of a cellular network using Orthogonal Frequency Division Multiplexing Access (OFDMA). On the other hand, by noting that perfect knowledge of the channel realization is a reasonable assumption to make for a static channel, \cite{MotahariGharanMaddahAliKhandani} analyzed a $M \times M$ Gaussian interference channel with real or complex static coefficients and employed a result from Diophantine approximation theory to prove the achievability of $\frac{MK}{2}$ degrees of freedom.
 
  It has been shown recently - for a frequency-selective single-input single-output (SISO) setup - by \cite{ThukralBolcskei} that the full spatial multiplexing gain of $\frac{K}{2}$ can be obtained even under conditions of limited feedback as long as the feedback rate exceeds $K(L-1) \log P$ bits per receiver, where $L$ is the number of taps in the channel between any pair of nodes and $P$ is the total power available with the transmitting sources. Their scheme involved a Grassmannian line quantization -  as in \cite{Mukkavilli} - of the channel vectors by the channel-aware receivers and their feedback using a limited number of bits broadcast to all the nodes in the network. The transmit and receiver beamforming vectors are then calculated at each node by treating these channel estimates as the actual channel realizations in the original IA procedure. While such an implementation would not align the interference perfectly, \cite{ThukralBolcskei} demonstrated that their scheme keeps the interference power bounded in the relevant-signal dimensions, thereby achieving the full multiplexing gain. Further, an analogous result is provided by the same authors in \cite{JatinPhDThesis} for the MISO network with the number of antennas at each transmitter being more than or equal to the number of users. Note that in such a scenario, interference alignment is \emph{not} needed, as zero forcing combined with beamforming suffices to attain the maximum sum degrees of freedom. 
     
     The sum degrees of freedom for the $2$-user multiple-input multiple-output (MIMO) interference channel - with ($M_1, M_2$) antennas at the two transmitters and ($N_1, N_2$) antennas at the two receivers - was carried out in \cite{JafarFakhereddin}, where the achievability depended solely on beamforming and zero-forcing techniques.  An extension to the $K$-user MIMO case with $M_t$ antennas at each transmitter and $M_r$ antennas at each receiver was provided in \cite{GouJafar}. The sum degrees of freedom was bounded within close lower and upper bounds which coincided whenever the ratio $\frac{\max \{ M_t, M_r \} }{\min \{ M_t,M_r \} }$ was an integer.  In particular, this means that the spatial multiplexing gain is precisely known for both the single-input multiple-output (SIMO) and the multiple-input single-output (MISO) cases.  The lower bounds obtained in \cite{GouJafar} depend on IA with perfect channel knowledge at each node. 
     
     In this paper, we explore all these multiple antenna cases, for a frequency-selective setup,  under the regime of limited feedback. In the SIMO case with $R$ antennas at each receiver, we find that as long as each receiver transmits no less than $N_f\ =\  K(RL - 1) \log P$ bits using quantization based on the {\em composite} Grassmann manifold, the nodes can utilize these channel estimates within the ambit of the IA scheme of \cite{GouJafar} and still achieve the complete spatial multiplexing gain.  The MISO case then follows immediately as a consequence of the reciprocity of alignment principle enunciated in  \cite{GomadamCadambeJafar}. Extending the same idea to the $K$-user MIMO case with $M_t$ antennas at each transmitter and $M_r$ antennas at each receiver, we find that our scheme of IA involving limited feedback attains the same degrees of freedom as the original IA scheme as long as each receiver uses no less than $N_f\ =\  {\min \{M_t,M_r \} }^2 K(RL - 1) \log P$ bits, where $R = \lfloor \frac{\max \{ M_t, M_r \} }{\min \{ M_t,M_r \} } \rfloor$.  In each case, a codebook over the composite Grassmann manifold is employed to jointly quantize the normalized channel directions.  We further generalize these results by proving that if each user transmitted a fraction of the afore-mentioned number of feedback bits, then the user can still attain a proportionate fraction of the degrees of freedom promised above.  In particular, if user $i$ fed back $\alpha_i \cdot N_f$ $( 0 < \alpha_i \leq 1)$ bits, then user $i$ achieves degrees of freedom equal to $\alpha$ times what that user would achieve with global perfect CSIT. Interestingly therefore, a slower scaling of feedback rate for any one user leads to commensurately fewer degrees of freedom for that user alone.
  
       This paper is organized into five sections. Section \ref{Introduction} introduces the topic and delineates the notational convention followed in this paper.  Section  \ref{SIMO_Section} computes the feedback scaling rate in the case of SIMO and MISO interference channels.  Section  \ref{MIMO_Section} explores the MIMO $M_t \times M_r$ case and shows that interference alignment with limited feedback can attain the best-known lower bound on the sum degrees of freedom for this network, which coincides with the actual sum degrees of freedom whenever $\frac{\max \{ M_t, M_r \} }{\min \{ M_t,M_r \} } $ is an integer.  Section \ref{RemarksDiscussion_Section}  evaluates the impact of a reduced feedback rate on the achievable degrees of freedom of the system.  This section also interprets our result in light of some earlier results in this direction and discusses insights drawn from our analysis. Section  \ref{Conclusion} concludes the paper. 
     
        Notation:  $\mathbb{R}$ and $\mathbb{C}$ represent the real and complex fields, respectively.  If $z \in  \mathbb{C}$, then $z^c$ represents its complex conjugate. The superscripts ${}^H$ and ${}^t$ represent the hermitian conjugate and the transpose of a matrix, respectively. Extending the conjugate notation to vectors, we denote as $v^c = [v_1^c, v_2^c, \ldots, v_n^c]^t$ when $v = [v_1, v_2, \ldots, v_n]^t$.  The square brackets $ [.] $ and the letter $l$ denote the time-slot index, usually varying from $0$ to $L-1$.  The circular brackets $ (.) $ and the letter $m$ denote the frequency-slot index, usually varying from $0$ to $N-1$. The symbols $\circ$ and $\otimes$ represent the Hadamard and Kronecker products of matrices, respectively. We use the term `$\mathrm{const}$' to denote any constant independent of the power $P$, whose value might change from equation to equation. This notational device greatly simplifies the appearance of many equations and enables us to concentrate on the relevant portions of the calculations. $\mathrm{CN}$ represents the circularly symmetric complex normal distribution. All logarithms in this paper are taken with respect to base $2$.

 \section{For Single-Input Multiple-Output (SIMO) Systems}\label{SIMO_Section}   
 
\subsection{System Model}
  
  There are $K$ single-antenna sources $S_1, S_2, \ldots, S_K$ with a message each for their respective single-antenna destinations $D_1, D_2,  \ldots, D_K$.  In our $K$-user SIMO interference channel model, we assume that each transmitter has one antenna and each receiver has $R$ antennas.   The $L$-tap response between the source $S_k$ and the destination $D_i$ is given by $h_{i,k}[l] \in \mathbb{C}^{R \times 1},\  l \in \{  0, 1, \ldots, L-1 \}$. We shall assume that these coefficients are drawn i.i.d from a continuous distribution such that these values are bounded with probability one and that the channel remains in the outage setting where the values $h_{i,k}[l]$ do not change during the transmission of the signal.   Let us define the matrix $T_{i,k}   \in \mathbb{C}^{L \times R}$ with the $L$  rows $h_{i,k}^t[0],   h_{i,k}^t[1],  \ldots, h_{i,k}^t[L-1]$ and denote its $R$ columns as $c_{i,k}[1],   c_{i,k}[2],  \ldots, c_{i,k}[R]$.  
\begin{equation*}
T_{i,k} \triangleq
  \begin{pmatrix}
h_{i,k}^t[0] \\
h_{i,k}^t[1] \\
\vdots \\
h_{i,k}^t[L-1]
\end{pmatrix}  
\triangleq   
  \begin{pmatrix}
c_{i,k}[1],   c_{i,k}[2],  \ldots, c_{i,k}[R]
\end{pmatrix} .
  \end{equation*}
  
The use of an OFDM-type cyclic signal model transforms this into $N$ parallel frequency flat channels, $h_{i,k}(r) \in \mathbb{C}^{R \times 1},\  r \in \{  0, 1, \ldots, N-1 \}$.
  The input-output relations are described by a set of $R$ equations, 
  \begin{equation}  y_i(r) = h_{i,i}^c(r) x_i(r)  +  \sum_{k \ne i} h_{i,k}^c(r) x_k(r) + z_i(r).  \end{equation}
  
  Here $y_i(r) , z_i(r) \in \mathbb{C}^{R \times 1}$ are  the channel output  and the i.i.d. noise on the $r$-th tone at the $i$-th receiver, $x_i(r) \in \mathbb{C}$ is the channel input on the $r$-th tone at the $k$-th transmitter and $h_{i,k}(r) \in \mathbb{C}^{R \times 1}$ is the channel vector from $S_k$ to $D_i$ on the same tone.  By defining the column vectors 
  \[ \overline{y}_i \triangleq [y_i(0), y_i(1), \ldots, y_i(N-1)] \in \mathbb{C}^{NR \times 1} , \] 
  \[ \overline{x}_i \triangleq [x_i(0), x_i(1), \ldots, x_i(N-1)] \in \mathbb{C}^{N \times 1} ,  \]
   \[ \overline{z}_i \triangleq [z_i(0), z_i(1), \ldots, z_i(N-1)] \sim CN(0, N_o I) \in \mathbb{C}^{NR \times 1},  \] and the block diagonal matrix 
  \[ \overline{H}_{i,k} \triangleq \mathrm{diag} \{ h_{i,k}^c(0), h_{i,k}^c(1), \ldots, h_{i,k}^c(N-1)  \} \in \mathbb{C}^{NR \times N} ,  \]
  we get the following equation 
 \begin{equation} \overline{y}_i = \overline{H}_{i,i} \overline{x}_i  +  \sum_{k \ne i} \overline{H}_{i,k}  \overline{x}_k + \overline{z}_i.   \end{equation}

Let us define the matrix $F_{i,k}\  \in \mathbb{C}^{N \times R}$ with the $N$ rows $ h_{i,k}^t(0), h_{i,k}^t(1), \ldots, h_{i,k}^t(N-1)$ and denote its $R$ columns as $h_{i,k}^{1}, h_{i,k}^{2}, \ldots, h_{i,k}^{R}$.  Note that $h_{i,k}^{m}$ is the DFT of $c_{i,k}[m]\  \forall m \in \{ 1,2, \ldots, R  \}$.
\begin{equation*}
F_{i,k} \triangleq
  \begin{pmatrix}
h_{i,k}^t(0) \\
h_{i,k}^t(1) \\
\vdots \\
h_{i,k}^t(N-1)
\end{pmatrix}  
\triangleq   
  \begin{pmatrix}
h_{i,k}^{1}, h_{i,k}^{2}, \ldots, h_{i,k}^{R}
\end{pmatrix} .
  \end{equation*}

 The receivers are assumed to have perfect knowledge of their respective channels, i.e.  each destination $D_i$ knows $T_{i,k}  \in \mathbb{C}^{L \times R}\  \forall k \in \{ 1,2, \ldots K \}$ perfectly.   As in the SISO case analyzed in \cite{ThukralBolcskei}, we assume there exist error-free dedicated broadcast links from the destinations to every other node in the network. During an initial channel feedback phase, the receiver broadcasts its channel state information using $N_f$ bits of feedback. This is followed by the data transmission phase.  The maximal rate of communication between $S_i$ and $D_i$ such that the probability of error is driven to zero as the block-length goes to infinity is denoted as $R_i$ with $R_{sum} \triangleq \sum_{i = 1}^K R_i$.    The sum degrees of freedom is defined as $d_{sum} \triangleq \lim_{P \rightarrow \infty} \frac{R_{sum}}{\log P},$ where $P$ is the total power constraint on the transmitting nodes. 

\subsection{Hamming Bound on the Composite Grassmann Manifold}

The complex Grassmann manifold $G_{n,k}$, viewed as a set, is  the collection of all the $k$-dimensional subspaces within a $n$-dimensional Euclidean space $\mathbb{C}^n$.  An alternate viewpoint, which is often useful, is to view it as an equivalence class of $n \times n$ unitary matrices defined in $G_{n,k} = U(n)/ (U(n-k) \times U(k))$.  A point on $G_{n,k}$ is then given by 
\begin{eqnarray*}
& [Q] & =  \\
 &{} &  \left \{    Q  \left (        \begin{array}{cc} 
                                                                                     Q_k & 0  \\ 
									 0 & Q_{n-k}  \end{array}                         \right ): Q_k \in U(k),  Q_{n-k} \in U(n-k)        \right \}  ,  
									    \end{eqnarray*}		
where $Q \in U(n)$. The study of the Grassmann manifold for wireless communication scenarios was boosted by the relation found between the manifold and non-coherent communication as reported in \cite{Zheng}. \cite{Mukkavilli} utilized Grassmannian vector quantization in their analysis of finite-rate feedback MIMO point-to-point links. This line of thought has been explored in great detail by many authors, as in \cite{MondalDuttaHeath} and \cite{LoveHeathStrohmer}. On the theoretical front, \cite{DaiJournal1}  computed the volume of a geodesic ball in $G_{n,k}$ and used the same to analyze MIMO systems under beamforming direction feedback.

Multidimensional Grassmann analysis (i.e. on $G_{n,k}$ when $k \ne 1$) cannot be directed in the current interference alignment scheme for a 3-user system, since if all that the users know is the subspace spanned by $\{  \frac{ h_{1,1} }{ || h_{1,1} || }, \frac{ h_{1,2} }{ || h_{1,2} || },   \frac{ h_{1,3} }{ || h_{1,3} || }    \}$, they would not be able to align their vectors so that they are separable at receiver one. A more precise characterization in terms of the actual channel directions is indispensable here for the accomplishment of the full spatial multiplexing gain. Such a representation is precisely what is provided by procedures on the composite Grassmann manifold.

In \cite{Dai} the composite Grassmann manifold $G_{n,k}^m$  is formed by taking the direct sum of $m$ copies of the Grassmann manifold $G_{n,k}$, i.e.
\[  G_{n,k}^m\   =\    \bigoplus_{m \mbox{ copies}}   G_{n,k}.  \]
On the Grassmann manifold, a commonly used distance metric is the chordal distance $d_c$. For the particular case of $k = 1$, it reduces to $d_c^2(v_1, v_2)\  \triangleq\  1 - | v_1^H v_2 |^2$. One can extend this distance to $G_{n,1}^m$ as follows:   If $P, Q\  \in\ G_{n,1}^m$, then $P\ =\  [p_1,\  \ldots,\ p_m]$, and $Q\ =\  [q_1,\  \ldots,\ q_m]$, where $p_i, q_i\  \in G_{n,1}\  \forall\  i\  \in \{ 1,2, \ldots, m  \}$.
\[ d^2(P, Q)\   \triangleq\   \sum_{i = 1}^m\  d_c^2(p_i, q_i).   \]  
In our analysis, we would need only $G_{RL,1}^{K}$; so we particularize the succeeding discussion in this subsection to this particular manifold.

One can define a ball of radius $\delta$ around a point $P \in G_{RL,1}^{K}$ as 
\[  B_P(\delta)\  =\   \{\  Q \in G_{RL,1}^{K}\  |\   d(P,Q) \leq \delta\     \}.  \]
Let $\mathrm{Vol}(A)$ denote the measure of the set $A$, with the measure being derived from the above defined distance metric.  Since $G_{RL,1}^{K}$ is an homogenous space, $\mathrm{Vol}(B_P(\delta))\  =\  \mathrm{Vol}(B_Q(\delta))\  \forall\  P,Q\  \in G_{RL,1}^{K}$. Hence, we can denote a ball of radius $\delta$ in the composite Grassmann manifold as $B(\delta)$ without any reference to its central point.  Based on our distance metric, let us construct a maximal packing of spheres on the composite Grassmann manifold $G_{RL,1}^{K}$ such that the minimum distance between the centers of any two spheres is more than $\delta$.  This is the precise analogue of the Grassmannian sphere-packing problem considered in  \cite{Conway}. Let us choose this maximal packing code as our quantization codebook ${\cal C}$.  The number of codewords possible in such a maximal packing is given by the well-known Hamming bound, as follows:
\[ |{\cal {C}}|\   \leq\     \frac{1}{ \mu (B(\delta))} .   \]

Here, $|{\cal {C}}|$ represents the number of codewords in our code ${\cal {C}}$, and $ \mu (B(\delta))$ represents the normalized volume of a geodesic ball in the composite Grassmann manifold, defined as $\mu (B(\delta))\  =\   \frac{\mathrm{Vol}(B(\delta))}{\mathrm{Vol}(G_{RL,1}^{K})} $.  We note that since the distances on the manifold are measured along geodesics (formed according to the choice of metric), the ball described below is formally called a \emph{geodesic} ball in the manifold.

A general result in \cite{Gray} states that for any valid distance metric, the normalized volume of the ball is given by the following expansion :  
\begin{equation}  \mu (B(\delta))\    =\   \mathrm{const}.\ \delta^{\dim G_{RL,1}^{K} }. (1 + o(\delta^2))        . \label{GrayFormula}\end{equation}
Here, $\dim G_{RL,1}^{K}$ is the real dimension of the composite Grassmann manifold and is given by $K \dim G_{RL,1} = 2K( RL - 1)$.  

Substituting $2^{N_f} \triangleq |{\cal {C}}|$, and noting that in the regime of increasing high $N_f$, the $o(\delta^2)$ term can be ignored, we write

\[    2^{N_f} \leq \frac{\mathrm{const}}{ \delta^{2K(R L - 1)}  } \Rightarrow \delta \leq \frac{ \mathrm{const} }{ 2^{ \frac{N_f}{2K(R L - 1)}}}  . \]

A realization $x \in G_{RL,1}^{K}$ shall be encoded using $N_f$ bits corresponding to the index of the codeword in ${\cal C}$ closest to it, i.e. the quantized version of $x$ shall be
\[  q(x)\   =\      \arg \min_{x_i \in {\cal C}}\   [d(x,x_i )] . \]
For our code, produced from a maximal packing on the composite Grassmann manifold, the maximum distortion experienced by a realization is bounded by the minimum distance of the code as follows:
\begin{equation}  \bigtriangleup_{\max} \leq \delta   \leq \frac{\mathrm{const}}{2^{\frac{N_f}{2K(RL - 1)}}}  , \label{MaxDistortionOnCGM}\end{equation}

\[ \mbox{where  }\ \bigtriangleup_{\max}\   \triangleq\    \max_{x  \in G_{RL,1}^{K}}  [d(x, q(x))]   .\]
Setting $\frac{1}{2^{\frac{N_f}{2K(RL - 1)}}}$ to be equal to $\frac{1}{\sqrt P}$, we get \begin{equation}
N_f = K \left (RL - 1 \right ) \log P \mbox{  bits.}
\label{FeedbackBitScalinginSIMOIA}\end{equation}

\subsection{Proposed Scheme}\label{IA_SIMO_ProposedScheme_Subsection}

The destination node $D_i$, knowing the matrix $T_{i,k}$ perfectly, forms a $RL$-length norm-one vector $v_{i,k}\ \triangleq \frac{ \mathrm{vec}(T_{i,k})}{ \|  \mathrm{vec}(T_{i,k})  \| }$. This calculation is performed for all $k \in \{ 1,2, \ldots, K \}$, creating a point on $G_{RL,1}^K$ denoted by $Q_i = [ v_{i,1}, v_{i,2}, \ldots, v_{i,K}]$. This is quantized over our $2^{N_f}$-level codebook ${\cal C}$ on the composite Grassmann $G_{RL,1}^K$ and is reconstructed by other nodes as $\hat{Q}_i = [ \hat{v}_{i,1}, \hat{v}_{i,2}, \ldots, \hat{v}_{i,K}]$.  The $RL$-length vector $\hat{v}_{i,k}$ is re-organized - similar to $T_{i,k}$ -  in form of a $L \times R$ matrix, 

\begin{equation*}
\hat{Q}_{i,k} \triangleq
  \begin{pmatrix}
\hat{w}_{i,k}^t[0] \\
\hat{w}_{i,k}^t[1] \\
\vdots \\
\hat{w}_{i,k}^t[L-1]
\end{pmatrix}  
\triangleq   
  \begin{pmatrix}
\hat{d}_{i,k}[1],   \hat{d}_{i,k}[2],  \ldots, \hat{d}_{i,k}[R]
\end{pmatrix} .
  \end{equation*}

The nodes zero-pad the $L$-length vectors $\hat{d}_{i,k}[m],\ m \in \{ 1,2, \ldots, R \}$ to $N$-length and take their DFTs to obtain $\tilde{d}_{i,k}[m]$. These are arranged as shown to form a $N \times R$ matrix, 

\begin{equation*}
\tilde{Q}_{i,k} \triangleq
  \begin{pmatrix}
\tilde{w}_{i,k}^t(0) \\
\tilde{w}_{i,k}^t(1) \\
\vdots \\
\tilde{w}_{i,k}^t(N-1)
\end{pmatrix}  
\triangleq   
  \begin{pmatrix}
\tilde{d}_{i,k}[1],   \tilde{d}_{i,k}[2],  \ldots, \tilde{d}_{i,k}[R]
\end{pmatrix} .
  \end{equation*}
The nodes also form the $RN \times N$ matrix $\tilde{W}_{i,k}$ as
 \[ \tilde{W}_{i,k} \triangleq \mathrm{diag} \{  \tilde{w}_{i,k}(0), \tilde{w}_{i,k}(1), \ldots, \tilde{w}_{i,k}(N-1) \} \in \mathbb{C}^{RN \times N}. \] 
 
  Now, we follow the  IA scheme  as in \cite{GouJafar} by treating the above $\tilde{W}_{i,k}$ as the actual $\overline{H}_{i,k}$ and utilize their interference alignment scheme of finding the direction vectors. Let $n \in \mathbb{N}$, 
  \begin{equation}  \Gamma \triangleq KR(K- R -1), \mbox{ and } \label{AuxVarGammaSIMOIA}\end{equation}
  \begin{equation} N \triangleq (R + 1)(n + 1)^{\Gamma} .   \label{NoOfTonesSIMOIA}\end{equation}
  We shall be coding over this $N$-symbols to achieve the following degrees of freedom : 
    \begin{equation}   d_i \triangleq \left \{     \begin{array}{ll} 
								R(n + 1)^{\Gamma} &  i \in \{ 1,2, \ldots, R+ 1\};\\ 
								R(n)^{\Gamma} &  i \in \{ R + 2, \ldots, K\}. \end{array}       \right.             
   \label{IndDegreesSIMOIA}\end{equation} 
 Now the transmitter and receiver pair $i$ choose $d_i$ different transmit and receive beamforming vectors, viz. $u_i^m \in \mathbb{C}^{RN \times 1}$ and $v_i^m \in \mathbb{C}^{N \times 1}$, respectively.  As long as $L$ scales sufficient fast with $\Gamma$ (cf. bandwidth scaling in \cite{GrokopTseYates}), one expects to find the vectors $u_i^m$ and $v_i^m$ satisfying the following properties. 

  \begin{eqnarray} | (u_i^m)^H \tilde{W}_{i,i}  (v_i^m)| \geq c > 0,\  \nonumber \\
  \forall i \in \{ 1,2, \ldots, K \},m \in \{1,2, \ldots, d_i \} ,  
\end{eqnarray}  
  
\begin{eqnarray} (u_i^m)^H \tilde{W}_{i,i}  (v_i^p)  = 0,\   \nonumber \\ 
 \forall i  \in \{ 1,2, \ldots, K \} \mbox{ and } \forall m \ne p \in  \{1,2, \ldots, d_i \} , 
\end{eqnarray}  
  
\begin{eqnarray} (u_i^m)^H \tilde{W}_{i,k}  (v_k^p)  = 0,\  \nonumber \\ 
\forall i \ne k \in \{ 1,2, \ldots, K \} \nonumber \\
 \mbox{ and } \forall m \in \{1,2, \ldots, d_i \} ,p \in \{1,2, \ldots, d_k \} .  
\end{eqnarray}

The source $S_k$ formulates $d_k$ independent symbols -- $x_k^1, \ldots, x_k^{d_k} \in \mathbb{C}$, which are sent along the directions $v_k^1, \ldots, v_k^{d_k} \in \mathbb{C}^{N \times 1}$ as follows :
  \begin{equation} 
    \overline{x}_k  \triangleq \sum_{m = 1}^{d_k} x_k^m v_k^m,   
  \end{equation} 
where $|| v_k^m || = 1$ and $ E (  |x_k^m|^2 ) = \frac{P}{K . d_k}$. Note that in the scheme of \cite{GouJafar}, \[ v_1^m\ =\ \cdots\ =\  v_{R + 1}^m,  \]  \[  \mbox{and           } v_{R + 2}^p\  =\  \cdots\  =\   v_K^p \]  for all valid values of $m$ and $p$.   We have implicitly assumed here that the number of users $K$ is greater than the number of receive antennas $R$ per node in the SIMO system. In the case of $K < R$, no interference alignment is necessary as mere zero-forcing would attain the maximal attainable $K$ degrees of freedom.

\subsection{Achievability Result}\label{IA_SIMO_Achievability_Subsection}

\vspace{2mm}
\begin{theorem}   The interference alignment scheme delineated above for a general $K$-user single-input multiple-output (SIMO) interference channel with one antenna at each transmitter and $R$ antennas at each receiver, where each pair of nodes in the network has a $L$-tap frequency selective channel between them,  achieves the full spatial multiplexing gain of $\frac{KR}{R+1}$ as long as each destination transmits more than $K(RL - 1) \log P$ bits of feedback, where $P$ represents the total power available with the transmitting nodes of the network. 
\label{SIMOTheoremDOFunderIA}\end{theorem}
\vspace{4mm}

\begin{proof}
We recall that using the Hamming bound on the composite Grassmann manifold $G_{RL, 1}^{K}$, one gets 
\begin{equation}  \bigtriangleup_{\max}   \leq \frac{\mathrm{const}}{2^{\frac{N_f}{2K(RL - 1)}}}  . \end{equation}
 Setting $ \frac{1}{2^{\frac{N_f}{2K(RL - 1)}}}$ as $\frac{1}{P}$, one obtained that $N_f = K(RL - 1) \log P$. Our aim is to show that this scaling of bits is sufficient to bound the interference terms in the rate expression. 
The destination $D_i$ projects the received signal $\overline{y}_i$ onto the $d_i$ directions given by $u_i^m \in \mathbb{C}^{RN \times 1}, m \in \{ 1,2,  \ldots, d_i\}$, 
\begin{eqnarray}
(u_i^m)^H \overline{y}_i & = &  (u_i^m)^H \overline{H}_{i,i} v_i^m x_i^m +  \sum_{p \ne m}  (u_i^m)^H \overline{H}_{i,i} v_i^p  x_i^p \nonumber \\
 & + &  \sum_{k \ne i} \sum_{p = 1}^{d_k}  (u_i^m)^H \overline{H}_{i,k} v_k^p  x_k^p  + (u_i^m)^H \overline{z}_i .
\end{eqnarray}
Let us define $1_R \triangleq [1,1, \ldots, 1]^t \in \mathbb{R}^{R \times 1}$,
 \[ \overline{h}_{i,k} \triangleq [ h_{i,k}^t(0),  h_{i,k}^t(1),  \ldots,  h_{i,k}^t(N-1) ] \in \mathbb{C}^{RN \times 1}, \]  and $\tilde{w}_{i,k} \triangleq [ \tilde{w}_{i,k}^t(0),  \tilde{w}_{i,k}^t(1),  \ldots,  \tilde{w}_{i,k}^t(N-1) ] \in \mathbb{C}^{RN \times 1}$.  

Let us define $(u_i^m)^c \circ ( v_k^p \otimes 1_R) $ as $b_{i,k}^{m,p} \in \mathbb{C}^{RN \times 1}$.  Then, we can denote $(u_i^m)^H \overline{H}_{i,k} v_k^p$ as $\overline{h}_{i,k}^H b_{i,k}^{m,p}$.
Using this notation, 

\begin{eqnarray}
(u_i^m)^H \overline{y}_i   & = &  \overline{h}_{i,i}^H b_{i,i}^{m,m} x_i^m  +  \sum_{p \ne m}  \overline{h}_{i,i}^H b_{i,i}^{m,p} x_i^p \nonumber \\
 & + &  \sum_{k \ne i} \sum_{p = 1}^{d_k}  \overline{h}_{i,k}^H b_{i,k}^{m,p} x_k^p  + (u_i^m)^H \overline{z}_i .
\end{eqnarray}
We again choose the input symbols $x_i^m$ to be i.i.d Gaussian and since the receiver knows $b_{i,i}^{m,m}$ and $  \overline{h}_{i,i}$, it can treat the other interference as noise to obtain a rate of 

\begin{equation}  R_i  \geq \frac{1}{N} \sum_{m = 1}^{d_i} \log \left ( 1 + \frac{ \frac{P}{Kd_i} |\overline{h}_{i,i}^H b_{i,i}^{m,m}   |^2 }{I_{i,1} + I_{i,2} + N_o} \right )  \mbox{,       with} \label{SIMOUserRateUnderCGIA}\end{equation}
\begin{equation} I_{i,1}  =  \sum_{p \ne m} \frac{P}{Kd_i} | \overline{h}_{i,i}^H b_{i,i}^{m,p} |^2 ,\mbox{        and} \label{IntFromOwnTxDiffBands}\end{equation}
\begin{equation}    I_{i,2}  =     \sum_{k \ne i} \sum_{p = 1}^{d_k}  \frac{P}{Kd_k} | \overline{h}_{i,k}^H b_{i,k}^{m,p} |^2    .    \label{IntFromOtherTx}\end{equation}

$I_{i,1}$ is the interference (treated as noise) caused by transmitter $i$ involving messages other than the one being currently decoded by receiver $i$.  $I_{i,2}$ is the interference caused by the transmitters other than transmitter $i$. 

  Our three conditions on the vectors $u_i^m$ and $v_i^m$ can also be re-written now as 
 \begin{equation} |\tilde{w}_{i,i}^H  b_{i,i}^{m,m}   |  \geq c > 0\   \forall i,m,    \label{IAmisoCONDone}\end{equation}
 \begin{equation}  \tilde{w}_{i,i}^H b_{i,i}^{m,p} = 0\ \forall i, m \ne p  ,     \label{IAmisoCONDtwo}\end{equation}
 \begin{equation}  \tilde{w}_{i,k}^H b_{i,k}^{m,p} = 0\ \forall k \ne i, \forall m,p .     \label{IAmisoCONDthree}\end{equation}
 
 We observe that $\overline{h}_{i,i}, \tilde{w}_{i,i}$ and $b_{i,i}^{m,p}$ are all $RN$-length vectors; and that 
 \begin{eqnarray*}
 \| \tilde{w}_{i,i} \|^2  & = & \sum_{n = 0}^{N -1}  \| \tilde{w}_{i,i}(n) \|^2 = \sum_{m = 1}^{R}  \| \tilde{d}_{i,i}[m] \|^2  \\
 & = & \sum_{m = 1}^{R}  \| \hat{d}_{i,i}[m] \|^2 = \| \hat{v}_{i,i} \|^2 = 1.
 \end{eqnarray*}
 Since, one can always append vectors to the orthogonal vectors $\tilde{w}_{i,i}$ and $b_{i,i}^{m,p}$ to form a basis for $\mathbb{C}^{RN}$, it follows that,
 \[  \left \|    \overline{h}_{i,i} \right \|^2  \geq   \left |  ( \overline{h}_{i,i})^H \tilde{w}_{i,i}    \right |^2  + \left |  ( \overline{h}_{i,i})^H \frac{b_{i,i}^{m,p}}{|| b_{i,i}^{m,p} ||}  \right |^2 . \]

We thus get,
\begin{eqnarray*}
& {} &  \frac{P}{Kd_i} | \overline{h}_{i,i}^H b_{i,i}^{m,p} |^2\  \\
& \leq &  \frac{P}{Kd_i}.\   \| b_{i,i}^{m,p}   \|^2  \left (      \left \|   \overline{h}_{i,i} \right \|^2      -    \left |  ( \overline{h}_{i,i})^H  \tilde{w}_{i,i}    \right |^2 \right ) \\
& \leq & \frac{P}{Kd_i}.\   \| b_{i,i}^{m,p}   \|^2   \| \overline{h}_{i,i}  \|^2  \left (      1     -    \left |  \frac{( \overline{h}_{i,i})^H}{\|  \overline{h}_{i,i} \|}  \tilde{w}_{i,i}    \right |^2 \right ).
\end{eqnarray*}
Now, note that
\begin{eqnarray*}
 \frac{( \overline{h}_{i,i})^H}{\|  \overline{h}_{i,i} \|}  \tilde{w}_{i,i} & = &  \sum_{j = 1}^{R}  \frac{(h_{i,i}^j)^H }{\|  \overline{h}_{i,i} \|}  \tilde{d}_{i,i}[j] \\
 & = & \sum_{m = 1}^{R}  \frac{(c_{i,i}[m])^H }{\|  \overline{h}_{i,i} \|}  \hat{d}_{i,i}[m] \\
& = & v_{i,i}^H \hat{v}_{i,i}.
\end{eqnarray*}
The last line follows from the observation that,
\begin{eqnarray*}
\| \overline{h}_{i,i} \|^2 & = &    \sum_{n = 0}^{N - 1} \| \overline{h}_{i,i}(n) \|^2 =   \sum_{m = 1}^{R} \| h_{i,i}^m \|^2 \\
& = &   \sum_{m = 1}^{R} \| c_{i,i}[m] \|^2 = \| \mathrm{vec}(T_{i,k})  \|^2.
\end{eqnarray*}
This leads to 
\begin{eqnarray*}
& {} &  \frac{P}{Kd_i} | \overline{h}_{i,i}^H b_{i,i}^{m,p} |^2\  \\
& \leq & \frac{P}{Kd_i}.\   \| b_{i,i}^{m,p}   \|^2   \| \overline{h}_{i,i}  \|^2  \left (      1     -    \left |  \frac{( \overline{h}_{i,i})^H}{\|  \overline{h}_{i,i} \|}  \tilde{w}_{i,i}    \right |^2 \right ) \\
& = &  \frac{P}{Kd_i}.\   \| b_{i,i}^{m,p}   \|^2   \| \overline{h}_{i,i}  \|^2   .\ (      1     -    |  v_{i,i}^H \hat{v}_{i,i}  |^2 ) \\
& = & \frac{P}{Kd_i}.\   \| b_{i,i}^{m,p}   \|^2   \| \overline{h}_{i,i}  \|^2  .\  d_c^2( v_{i,i},\  \hat{v}_{i,i}) \\
& \leq & \frac{P}{Kd_i}.\   \| b_{i,i}^{m,p}   \|^2   \| \overline{h}_{i,i}  \|^2  .\  \sum_{k = 1}^K d_c^2( v_{i,k},\  \hat{v}_{i,k}) \\
& = & \frac{P}{Kd_i}.\   \| b_{i,i}^{m,p}   \|^2   \| \overline{h}_{i,i}  \|^2  .\  d^2( Q_i,\  \hat{Q}_i) \\
              &  \leq &  P .\  \mathrm{const} .\ \bigtriangleup_{\max}^2    \\
              &  \leq &  P.\  \frac{\mathrm{const}}{2^{\frac{N_f}{K(RL - 1)}}} \\
              &  =  &  P .\  \mathrm{const} .\ \frac{1}{P} \\
              &  =  &  \mathrm{const}.
\end{eqnarray*}

Hence, $I_{i,1}$ is bounded.   A similar argument using $\frac{ \sqrt P  }{ \sqrt {Kd_k}}   |  \overline{h}_{i,k}^H  b_{i,k}^{m,p} |$  shows that $I_{i,2}$ is also bounded independent of power $P$.  Further, as $P \rightarrow \infty$ (and the number of feedback bits $N_f \rightarrow \infty$), the quantization error becomes vanishingly small by equation (\ref{MaxDistortionOnCGM}).  This implies that, as in \cite{ThukralBolcskei}, that 
\[ \tilde{w}_{i,i} \rightarrow \frac{\overline{h}_{i,i}}{\|  \overline{h}_{i,i} \|}\  \Rightarrow\  | \overline{h}_{i,i}^H b_{i,i}^{m,m} |  > 0. \]

Now, the equation (\ref{SIMOUserRateUnderCGIA}) yields the following result:
\begin{eqnarray*}
d_{sum} & = & \lim_{P \rightarrow \infty} \frac{R_{sum}}{ \log P} \\
               & \geq &     \sum_{i = 1}^{K} \sum_{m = 1}^{d_i} \lim_{P \rightarrow \infty} \frac{ \log \left ( 1 + \frac{ \frac{P}{Kd_i} |\overline{h}_{i,i}^H b_{i,i}^{m,m}   |^2 }{I_{i,1} + I_{i,2} + N_o} \right ) }{N \log P}  \\
               & = &  \frac{ \sum_{i = 1}^{K} d_i}{N} \\
               & = & \frac{(R + 1) R (n  + 1)^{\Gamma}  +  (K - R - 1) R n^{\Gamma}}{(R + 1) (n  + 1)^{\Gamma}}
\end{eqnarray*}
The last step above follows from equations (\ref{AuxVarGammaSIMOIA}), (\ref{NoOfTonesSIMOIA}) and (\ref{IndDegreesSIMOIA}).
Taking the supremum over all values of the auxiliary parameter $n$, we obtain the desired result that $d_{sum} = \frac{RK}{R + 1}$, which is the same as was obtained in the case of perfect channel state information at all the nodes in \cite{GouJafar}. 

\end{proof}

The MISO case follows immediately by application of the reciprocity of alignment scheme discussed in \cite{GomadamCadambeJafar}.

\section{For Multiple-Input Multiple-Output (MIMO) Systems}\label{MIMO_Section}

The precise degrees of freedom for the general $M_t \times M_r$ $K$-user interference channel have not yet been quantified. Gou and Jafar \cite{GouJafar} established lower and upper bounds which coincide when $R =  \lfloor \frac{ \max \{ M_t,M_r \}  }{ \min \{ M_t,M_r \} } \rfloor$ is an integer.  We denote these as $d_{sum}^{UB}$ and $d_{sum}^{LB}$ below. 
    \begin{equation*}   d_{sum}^{UB}  = \left \{     \begin{array}{ll} 
								K. \min \{ M_t,M_r \} & K \leq R;\\ 
								\frac{\max \{ M_t,M_r \}}{ R + 1}.K &  K>R. \end{array}       \right.             
   \end{equation*} 
In the case of $K \leq R$, the $d_{sum}^{UB} $ is obtained by noting that in the event of facing no interference, the $d_{sum}$ is upper bounded by $K$ times the degrees of freedom of the single-user MIMO channel as calculated in \cite{Telatar}.  In the case of $K > R$, a tighter upper bound is obtained by first reducing the $K$-user channel to different $R+1$-user configurations; then by making the first $R$ of the transmitters and the last $R$ of the receivers cooperate; and finally by invoking the results presented in \cite{JafarFakhereddin} on the $2$-user case.

  The lower bounds are as follows:
    \begin{equation*}   d_{sum}^{LB}  = \left \{     \begin{array}{ll} 
								K. \min \{ M_t,M_r \} & K \leq R;\\ 
								\frac{R}{ R + 1}. \min \{ M_t,M_r \} K &  K>R. \end{array}       \right.             
   \end{equation*} 
The case of $K \leq R$ is handled using just beamforming and zero forcing. The case of $K > R$ is proved using interference alignment with perfect channel knowledge.  We show that an appropriate scaling of the feedback bits suffices to achieve this spatial multiplexing gain. 

\vspace{4mm}
\begin{theorem}
In a general $K$-user interference channel where each transmitter has $M_t$ antennas, each receiver has $M_r$ antennas, $K > R =  \lfloor \frac{ \max \{ M_t,M_r \}  }{ \min \{ M_t,M_r \} } \rfloor $, and each pair of nodes in the network has a $L$-tap frequency selective channel between them, an interference alignment scheme under limited feedback achieves the same degrees of freedom as the interference alignment scheme with perfect channel state information as long as the receiver employs more than ${\min \{M_t,M_r\} }^2 K(RL - 1) \log P$ bits of feedback, where $P$ is the total power available with the transmitting nodes of the network. 
\label{MIMOTheoremDOFunderIA}\end{theorem}
\vspace{4mm}

\begin{proof}
Due to the reciprocity of alignment idea \cite{GomadamCadambeJafar}, we can concentrate our attention on the case of $M_t \leq M_r$ without any loss of generality.  From each receiver node, we discard $M_r - RM_t$ antennas. Now, by restricting the cooperation allowed amongst the transmitters and the receivers, we can treat this $K$-user $M_t \times RM_t$ channel as a SIMO $KM_t$-user $1 \times R$ channel. As long as each user - utilizing composite Grassmann quantization - sends $(KM_t). (RL - 1). \log P$ bits of feedback, we know from the preceding section that a spatial muliplexing gain of $\frac{KM_tR}{R + 1}$ can be achieved. This value of $\frac{KM_tR}{R + 1}$ matches with the desired inner bound achieved with perfect channel knowledge in \cite{GouJafar}.  Combining back the receivers to get the $K$-user channel, we conclude that each node needs to transmit at least  $M_t^2K(RL - 1)$ bits of feedback.
\end{proof}

\section{Remarks and Discussion}\label{RemarksDiscussion_Section}

\subsection{Impact of smaller feedback scaling rate}

Through theorems \ref{SIMOTheoremDOFunderIA} and \ref{MIMOTheoremDOFunderIA}, we have established the achievement of the maximal sum degrees of freedom ($d_{sum} \triangleq \frac{\sum_{i =1}^{K} d_i}{N}$), when the feedback by each user scales as some $N_f$ bits.  However, in an interference network, it may not be possible on part of each user to scale its feedback bits at this uniform rate. Obversely, it would be desirable to have a tradeoff between the provision of a flexible feedback rate for an individual user and the degrees of freedom obtained thereby, with the system-level strategy being optimized (as in Sections \ref{SIMO_Section} and  \ref{MIMO_Section}) for achieving the Pareto-optimal point maximizing the sum degrees of freedom of the network. This tradeoff provides additional flexibility to the system designer in terms of allocation of resources for feedback and obtainable rates is given by the theorem below.  

We denote the degrees of freedom achieved by user $i$ as $ \tilde{d}_i \triangleq \frac{d_i}{N}$, where $d_i \triangleq \lim_{P \to \infty} \frac{R_i}{\log P}$ and $R_i$ is the rate achieved by user $i$ over $N$ blocks.  

\vspace{2mm}
\begin{theorem}
If the feedback rate employed by receiver $i$ in the interference channel (both the SIMO/MISO model of Section \ref{SIMO_Section}  and the MIMO model of Section \ref{MIMO_Section}) scales as $\alpha_i . N_f$, for some $0\ <\  \alpha_i\  \leq 1$, then user $i$  can achieve $\alpha_i. \tilde{d}_i$ degrees of freedom by using interference alignment. 
\end{theorem}
\vspace{2mm}

\begin{proof}
We analyze below for a $K$-user SIMO $1 \times R$ channel; the corresponding MIMO results follow in a similar manner.  Intuitively, the quantization error is proportional to $P^{1 - \alpha_i}$ and this acts as a principal component of the interference faced by user $i$, leading to a reduction of $(1 - \alpha_i) \tilde{d}_i$ degrees of freedom from user $i$. 

As in equation (\ref{LBonUserRateUnderIA}), the rate achieved by user $i$ can be lower bounded as 
\begin{eqnarray*}
       R_i & \geq &   \frac{1}{N} \sum_{m = 1}^{d_i} \log \left ( 1 +  \frac{P}{Kd_i} |\overline{h}_{i,i}^H b_{i,i}^{m,m}   |^2  \right ) \\
       & - &  \frac{1}{N} \sum_{m = 1}^{d_i} \log \left ( I_{i,1} + I_{i,2} + N_o \right ) . 
\end{eqnarray*}
The interference terms can be expressed as
\begin{eqnarray*}
& &  I_{i,1} +  I_{i,2} \\
&  =  & \sum_{p \ne m} \frac{P}{Kd_i} | \overline{h}_{i,i}^H b_{i,i}^{m,p} |^2  + \sum_{k \ne i} \sum_{p = 1}^{d_k}  \frac{P}{Kd_k} | \overline{h}_{i,k}^H b_{i,k}^{m,p} |^2    \\
& \leq & \sum_{p \ne m}  \frac{P}{Kd_i}.\   \| b_{i,i}^{m,p}   \|^2   \| \overline{h}_{i,i}  \|^2  .\  d_c^2( v_{i,i},\  \hat{v}_{i,i}) \\
& + &  \sum_{k \ne i} \sum_{p = 1}^{d_k} \frac{P}{Kd_k}.\   \| b_{i,k}^{m,p}   \|^2   \| \overline{h}_{i,k}  \|^2  .\  d_c^2( v_{i,k},\  \hat{v}_{i,k}) .
\end{eqnarray*}
If we denote $b_{max}^2\  \triangleq  \max_{i,k,m,p} \| b_{i,k}^{m,p} \|^2$, and $h_{max}^2\  \triangleq  \max_{i,k} \|  \overline{h}_{i,k} \|^2$, then
\begin{eqnarray*}
& &  I_{i,1} +  I_{i,2} \\
& \leq & b_{max}^2.\ h_{max}^2.\ \mathrm{const}.\   P.\   \sum_{k = 1}^K d_c^2( v_{i,k},\  \hat{v}_{i,k}) \\
& = & P.\  \mathrm{const}.\     d^2( Q_i,\  \hat{Q}_i) \\
&  \leq &  P .\  \mathrm{const} .\ \bigtriangleup_{\max, i}^2    .
\end{eqnarray*}

Here, $\bigtriangleup_{\max, i}$ is the maximum quantization error possible while using the codebook of receiver $i$ with $2^{\alpha_i N_f}$ representations on $G_{RL, 1}^{K}$.  As in equation (\ref{MaxDistortionOnCGM}), we can claim that  
\begin{eqnarray*}
&& \bigtriangleup_{\max, i}\   \leq\  \frac{\mathrm{const}}{2^{\frac{\alpha_i N_f}{2K(RL - 1)}}} \\
& \Rightarrow &  \bigtriangleup_{\max, i}^2\  \leq\   \frac{\mathrm{const}}{2^{\frac{\alpha_i  K(RL - 1) \log P}{K(RL - 1)}}}\  =\  \frac{\mathrm{const}}{P^{\alpha_i}}
\end{eqnarray*}
This leads to,
\begin{eqnarray*}
&&  I_{i,1} +  I_{i,2}\  \leq\   \mathrm{const}. P^{1 - \alpha_i} \\
 && \Rightarrow MG(R_i)\  \geq\  \frac{1}{N} (d_i - (1 - \alpha_i) d_i) = \frac{\alpha_i d_i}{N} = \alpha_i \tilde{d}_i,
\end{eqnarray*}
where $MG(x)$ denotes the multiplexing gain of the quantity $x$ defined as $MG(x)\ =\ \lim_{P \to \infty} \frac{x}{\log P}$.
\end{proof}

In particular, if $\alpha_1\ =\ \alpha_2\ =\  \ldots\ \alpha_K\ = \alpha$ (say) , then $d_{sum}\  =\  \alpha \sum_{i = 1}^K \frac{d_i}{N} = \alpha . \frac{RK}{K + 1}$ in a $K$-user SIMO $1 \times R$ channel. A similar result would also hold for the MIMO $M_t \times M_r$ channel.  Also, note that while the rate achieved by user $i$ would depend upon $\{ \alpha_j \}_{j = 1}^{K}$, the degrees of freedom depends only on $\alpha_i$ as long as the other $\alpha_j $'s are positive.

For the case of $M$-user Rayleigh faded broadcast channel with $M$ antennas at the transmitter, \cite{Jindal} had analyzed the specific strategy of zero-forcing (ZF) precoding with random vector quantization or RVQ-generated quantization codebooks to show that a feedback of $\alpha$ times the optimal feedback rate leads to $\alpha$ times the maximum sum degrees of freedom.  The above theorem presents a counterpart to that result for the $K$-user MIMO interference channel with several differences.  We do not restrict the number of users to be equal to the number of antennas.  The fading distribution is more general. No constraint is placed on the quantization codebook.  Even when formulated using pseudo-beamforming vectors $b_{i,k}^{m,p}$, interference alignment differs widely from the simple ZF precoding schemes with the $b_{i,k}^{m,p}$ 's arising from involved functions of the channel matrices. 

\subsection{Shrinkage of `radius of uncertainty'}

The work in \cite{CaireJindalShamai}  analyzed the operation of a multi-antenna broadcast channel with the transmitter basing its beamforming vectors on imperfect channel estimates received from the receivers. Particularizing to the two-user case, they considered a scenario where the norm-one channel direction to user one is fixed at $h_1$ and the direction vector to user two can be either $h_{2a}$ or $h_{2b}$. Irrespective of the transmission strategy adopted at the transmitter,  \cite{CaireJindalShamai} showed that the square of the chordal distance between $h_{2a}$ and $h_{2b}$ must shrink as $O(P^{-1})$, i.e  $1 - | h_{2a}^H h_{2b}|^2 = O(P^{-1})$, for the system to achieve the same degrees of freedom as the perfect CSIT case. The achievability for the same was shown via a simple beamforming scheme.

An analogous formulation of our problem would be to note that the rate achieved by user $i$, as expressed in equation (\ref{SIMOUserRateUnderCGIA}), is 
\begin{eqnarray}
R_i & \geq &  I (X_i; Y_i\  |\   \{ \hat{Q_i} \}_{i}^{K}) \nonumber\\
      & \geq &  \frac{1}{N} \sum_{m = 1}^{d_i} \log \left ( 1 + \frac{ \frac{P}{Kd_i} |\overline{h}_{i,i}^H b_{i,i}^{m,m}   |^2 }{I_{i,1} + I_{i,2} + N_o} \right ) \nonumber\\
       & \geq &   \frac{1}{N} \sum_{m = 1}^{d_i} \log \left ( 1 +  \frac{P}{Kd_i} |\overline{h}_{i,i}^H b_{i,i}^{m,m}   |^2  \right ) \nonumber\\
       & - &  \frac{1}{N} \sum_{m = 1}^{d_i} \log \left ( I_{i,1} + I_{i,2} + N_o \right ) . 
\label{LBonUserRateUnderIA}\end{eqnarray}
Defining the multiplexing gain of a quantity $x$ as $\lim_{P \to \infty} \frac{x}{\log P}$, we note that the multiplexing gain of the first term equals the same degrees of freedom as available for the user in this channel in the perfect CSIT case. Further, we can show that the multiplexing gain of the second term is zero if both $I_{i,1} $ and $I_{i,2}$ are shown to be constant independent of $P$. This condition is satisfied, as seen from our calculations in Section \ref{SIMO_Section}.\ref{IA_SIMO_Achievability_Subsection}, if $\bigtriangleup_{\max}^2 \leq \frac{1}{P}$, which leads to $N_f = K \left (RL - 1 \right ) \log P$ bits in equation (\ref{FeedbackBitScalinginSIMOIA}). From Section \ref{SIMO_Section}.\ref{IA_SIMO_ProposedScheme_Subsection}, we note that $\bigtriangleup_{\max}^2 = \max_{ \hat{Q}_i \in {\cal{C}}} d^2(Q_i, \hat{Q}_i)$, where $Q_i \in G_{RL,1}^K$ and ${\cal{C}}$ is the code of cardinality $2^{N_f}$ used to quantize $Q_i$. Equivalently, one can re-write the necessary condition as  
\begin{eqnarray}
& {} & d^2(Q_i, \hat{Q}_i)\  \leq \frac{1}{P} \nonumber \\
& \Rightarrow &    d_c^2(v_{i,k}, \hat{v}_{i,k})\ \leq \frac{1}{P}\  \forall\ k \in \{ 1,2, \ldots, K \} .
\label{RadiusOfUncertaintyShrinkage}\end{eqnarray}

Recall that $v_{i,k}$ is the norm-one estimate of the channel vector; and $\hat{v}_{i,k}$ is its channel estimate provided by the receiver. As in the sufficiency part of the proof in \cite{CaireJindalShamai}, the $\hat{v}_{i,k}$  were treated as the actual channel directions in the formulation of the beamforming vectors ($u_i^m$ and $v_k^p$) by the nodes. In our analysis, the role corresponding to the beamforming vectors of the broadcast channel are performed by the pseudo-beamforming vectors $b_{i,k}^{m,p}$ formed by the Hadamard product of the transmit ($v_k^p$) and receive ($u_i^m$) beamforming vectors. Our results interpreted in this manner act as a counterpart for the case of interference channels to the sufficiency results in \cite{CaireJindalShamai} obtained for the broadcast channel.  The complicated dependence of the pseudo-beamforming vectors $b_{i,k}^{m,p}$ on the fedback channel information precludes finding a direct counterpart to the necessity part of their proof.

Note that in our case, $v_{i,k}$ is uniformly distributed over a space $G_{RL, 1}$ with uncountably infinite points, i.e. it  lies in a \emph{ball of uncertainty} around the estimate $\hat{v}_{i,k}$. Equation (\ref{RadiusOfUncertaintyShrinkage}) then shows that the shrinkage of its  \emph{radius of uncertainty} as $O(P^{-1})$ suffices to attain the same degrees of freedom as the perfect CSIT case.

\subsection{Remarks}

\begin{itemize}
\item The connection of the pre-log factor in the feedback scaling answer to the dimension of the manifold over which the quantization happens is intriguing.  In schemes involving beamforming feedback of a single vector of $M$-length, the dimension of this norm-one vector is $2(M - 1)$. Correspondingly, we find that $(M - 1) \log P$ bits of feedback suffices to obtain ideal-like performance.  In the interference alignment scheme explored above in the SIMO and MISO cases, we obtain exactly $\frac{1}{2} \dim G_{RL, 1}^{K}$ as the pre-log factor. Note that the real dimension of the composite Grassmann manifold $G_{RL, 1}^{K}$ is $2K(RL - 1)$. This line of thought is further explored in Appendix \ref{PrelogFactorFeedbackRate}.

%\item  When the number of users in the MIMO interference channel is exactly two and perfect channel knowledge is available at all the nodes, the maximal degrees of freedom were precisely determined in \cite{JafarFakhereddin}. Exploring the case when the receivers feedback information about the channel using a finite number of bits, we obtain the following result in Appendix \ref{2UserMIMOIntFdBckScaling}.

%\vspace{2mm}
%\begin{corollary}
%In a two-user MIMO interference channel with $M_1, M_2$ antennas at the transmitters, and $N_1, N_2$ antennas at the receivers, the complete sum degrees of freedom can be achieved as long as the two receivers are able to broadcast   $\min \{  N_1, M_2 \} . (M_2 - 1) \log P$ and $\min \{  N_2, M_1 \} . (M_1 - 1) \log P$  bits, respectively,  of feedback. 
%\end{corollary}
%\vspace{2mm}

\item Since our analysis in Section \ref{SIMO_Section}.\ref{IA_SIMO_Achievability_Subsection} only required the first-order  term in the volume expansion  of a ball in $G_{n,1}^K$, the expansion obtained in equation (\ref{GrayFormula})  was sufficient.  The structure of $G_{n,1}^K$, however, permits the evaluation of the precise volume of the ball as follows. 

\vspace{2mm}
\begin{theorem}
The normalized volume of a ball of radius $\delta$ in the composite Grassmann manifold $G_{n,1}^K$ is given by 
\[ \mu(B(\delta))\ =\   \frac{\Gamma^{K}(n)}{\Gamma(K (n - 1) + 1)}    \delta^{2K(n - 1) }   . \]
\end{theorem}
\vspace{2mm}

The proof is provided in Appendix \ref{PreciseBallVolCompGrassmann}

%\item  While this paper computes a sufficient scaling of feedback bits for the interference alignment scheme to attain the maximum multiplexing gain, it might or might not be the minimal scaling that is necessary. Such results on minimal scaling required are known for other scenarios as in \cite{Jindal}, which computes the same as $(M - 1 ) \log P$ bits for the case of Gaussian broadcast channels with $M > 1$ antennas at the transmitter and $M$ single-antenna receivers with the channel directions being quantized using the Random Vector Quantization strategy. 

\item It would be interesting to analyze interference alignment schemes under finite-rate feedback for time-selective channels. This paper's approach as also the analysis earlier by \cite{ThukralBolcskei} cannot be directly applied to time-selective channels without violating the non-causality requirement on channel knowledge at the receivers. 

\end{itemize}

\section{Conclusion}\label{Conclusion}
Multi-user interference channels with single or multiple antennas at each node are analyzed in a frequency-selective setup, wherein the receivers with knowledge of their respective channels quantize the channel directions using a code-book on the Composite Grassmann manifold and broadcast them to all other nodes at a rate that scales sufficiently fast with the power constraint on the nodes.  It is shown that an interference alignment scheme based on treating these channel estimates as being perfect is sufficient to attain the same sum degrees of freedom as the interference alignment implementation utilizing perfect channel state information at all the nodes. We also demonstrate a continuous tradeoff whereby an individual user can opt for a slower scaling of feedback bits and obtain proportionally lower degrees of freedom. 

\appendix 

\section{Pre-log Factor in Feedback Scaling Rate}\label{PrelogFactorFeedbackRate}

In this appendix, we give an intuitive explanation for why the pre-log factor in the feedback scaling rate required for emulating perfect CSI performance involves a term corresponding to the dimension of a manifold.  The performance measure seems to be immaterial here, viz. it can be the maximal spatial multiplexing gain, the capacity of a channel or the probability of error incurred in a scheme. One first isolates that channel parameter the knowledge of which enables the scheme (using operations at both the transmitter and the receiver) to attain its `ideal' value.  This channel parameter can be, for example, 
\begin{itemize}
\item a single beamforming vector $h \in \mathbb{C}^{L \times 1},\ \| h \|  = 1 \Rightarrow  \mathrm{dimension} = L - 1, $
\item an input covariance matrix $Q \in \mathbb{C}^{n \times n},\  \mathrm{Tr}\ Q \leq \rho \Rightarrow  \mathrm{dimension} = n^2, $
\item a set of directions $W \in G_{RL,1}^{K}\  \Rightarrow  \mathrm{dimension} = 2RK(L - 1). $
\end{itemize}
Let us denote this channel parameter as $q$, and denote its substratum on which it takes values as the manifold $M$. If the real dimension of $M$ is given by $\dim M = N$, then we know that around each point on the manifold, we can establish a local coordinate system with $N$ variables. For analysis, consider such a system around the optimal value of $q$  denoted as $q_{opt}$. If the receiver only has knowledge of $q$, it can decide to feedback this information to the transmitter using a finite number, say $N_f$, of bits.  By following the argument we made in Section \ref{SIMO_Section},  one can bound the maximum distortion sufferable by a point under quantization using a $2^{N_f}$-level codebook on the manifold as:
\[  \bigtriangleup_{\max}\   \leq\ \frac{\mathrm{const}}{2^{\frac{N_f}{N}}} . \]
Here, the codebook is the one that solves the packing problem of spheres for a given code minimum distance for the given manifold $M$. The bound then follows using the Hamming upper bound and noting a general formula for ball volumes in a general manifold. We note that if $q$ varies uniformly on the manifold $M$, then the distortion-rate function - which is the minimum distortion over the choice of all the codes - also varies as the inverse of $2^{\frac{N_f}{N}}$, as seen in \cite{OurAllertonPaper2008} and \cite{DaiJournal1}.

The receiver in attempting to convey $q_{opt}$ using $N_f$ bits causes an error,  and the transmitter is able to reconstruct the same as some $\hat{q}$. In the regime of high $N_f$, we can take $\hat{q}$ within the same local coordinate system that covers $q_{opt}$. Isolating the key part of the performance measure affected by the inaccuracy in $q$ as $f(q)$, we get an appropriate Taylor expansion:
\[ f(\hat{q})  \approx f(q_{opt}) +  (\nabla f)^t (\triangle q). \]
$\nabla f$ represents the gradient of the function $f$ and $\triangle q$ represents the displacement of $\hat{q}$ from $q_{opt}$. Note that this is a coarse approximation chosen primarily for illustration; to make it rigorous, one would have to choose a normal coordinate system based at $q_{opt}$ and calculate distances along radial geodesics emanating from it. 

Our intention is often to bound the difference between $f(\hat{q})$ and $f(q_{opt})$ by a constant independent of $P$; as we expect to see both $f(\hat{q})$ and $f(q_{opt})$ increase monotonically with $P$.  The variation of the $(\nabla f)$ with $P$ determines the behavior required of the $\triangle q$ term. If $(\nabla f)$ varies as $\sqrt P$, then we need $\triangle q$ to vary as $\frac{1}{\sqrt P}$. This is ensured by 
\begin{eqnarray*}
\| &  \triangle q &  \|^2\  \leq\ \frac{1}{P}\ \Rightarrow\ \bigtriangleup_{\max}^2\ \leq\ \frac{1}{P} \\
   & \Rightarrow & \frac{\mathrm{const}}{2^{\frac{N_f}{N}}}\  \leq\  \frac{1}{P}\ \Rightarrow\  N_f\ \approx\  \frac{N}{2} \log P \mbox{  as $P$ increases.}
\end{eqnarray*}
This provides an intuitive reason for our results on the feedback scaling rate in this paper as well as other similar results in \cite{ThukralBolcskei} and \cite{Jindal}.

\section{Precise Ball Volume in the Composite Grassmann Manifold}\label{PreciseBallVolCompGrassmann}

We are interested in finding the precise expression for the normalized volume of a ball in the Composite Grassmann manifold $G^K_{n,1}$.  For the general case of $G_{n,p}$, $p \in \{ 1,2, \ldots, n \}$, \cite{DaiJournal1} computed the first two terms in a series expansion for the normalized volume of a ball. The calculation has been extended in \cite{Dai} to the case of the Composite Grassmann $G^K_{n,p}$. In the special case of $G_{n,1}$, the ball volume expression, somewhat surprisingly, reduces to a single term, as is extracted below from a simple manipulation of the results in \cite{Mukkavilli}.  The result of \cite{DaiJournal1} for the case of general $p$ does not reduce by setting $p = 1$ to the expression in \cite{Mukkavilli} due to a mistake in their Corollary 1, wherein they claim the equivalence of the series expansion to its first term by showing merely that the second term in the expansion vanishes in certain cases. 

The paper \cite{Mukkavilli} considers a spherical cap defined by
\[ S_i(\gamma) =  \{   h \in \mathbb{C}^t\  |\  \| h \|^2 = \gamma,\   |< h, C_i > |^2  \geq \gamma_0    \}. \]  Here $C_i$ is an unit norm vector.  The area of this spherical cap is then calculated as 
\[ A(S_i(\gamma))  = \frac{2 \pi^t \sqrt{\gamma} (\gamma - \gamma_0)^{t  - 1} }{(t  - 1)!} . \]
By setting $\gamma = 1$, $t = n$, $\delta = \sqrt{1 - \gamma_0}$, the spherical cap can be seen to be a ball in $G_{n,1}$ of radius $\delta$,
\[ S_i(\gamma) =  \{   h \in \mathbb{C}^n\  |\  \| h \|^2 = 1,\  d_c^2(h, C_i )  \leq \delta^2   \}.  \]
The volume of the ball of radius $\delta$ is then seen to reduce to 
\[  \mathrm{Vol}(B(\delta)) = \frac{2 \pi^n \delta^{2(n-1)}}{ (n - 1)!}  . \]  Under the chordal distance metric, \cite{Dai} provides the volume of the entire Grassmann manifold $G_{n,1}$ as $\frac{2 \pi^n}{(n - 1)!}$.  This gives the normalized volume of the ball in $G_{n,1}$ as
\[   \mu(B(\delta)) =  \delta^{2(n-1)},   \]
as compared to the expression in \cite{DaiJournal1} which only yields $\mu(B(\delta)) =  \delta^{2(n-1)} ( 1 + 0. \delta^2  + o(\delta^2))$.

To extend the exact result to the Composite Grassmann case, we fix a point $P \in G^K_{n,1}$, i.e. $P \triangleq [P_1, P_2, \ldots, P_K],\  P_i \in G_{n,1} \forall i \in \{  1,2, \ldots, K\}$.  We vary a random variable $Q \triangleq [Q_1, Q_2, \ldots, Q_K]$ uniformly on $G^K_{n,1}$, viz. $Q_i \sim \mathrm{Unif}(G_{n,1})\ \forall i \in \{  1,2, \ldots, K\}$.  By defining  $K$ random variables $X_i \triangleq d_c^2(P_i, Q_i)$, we note that these are independent and identically distributed with the cumulative distribution function 
\[  F_{X_i}(x)  =  \Pr \{X_i  \leq x \}  =  \mu(B(\sqrt{x})) = x^{n -1} .  \]
Since the random variable $X_i$ measures the square of the cosine of the angle between two vectors, it is bounded between $0$ and $1$.  We also get the probability density function of $X_i$ as
\[   f_{X_i}(x)  =  (n - 1).x^{n - 2}. 1_{[0,1]} , \]
with $1_{[0,1]}$ being the standard indicator function for the interval $[0,1]$.

Let us define $U \triangleq \sum_{i = 1}^K X_i$ and we claim that the density of $U$ is given by 
\[  f_U(x) =  (n - 1)^K  x^{K(n-1) - 1} \frac{\Gamma^K(n - 1)}{\Gamma(K(n - 1))}  1_{[0,1]} . \]   This can be proved through induction.  Assume that statement is true for $\sum_{i = 1}^{K - 1}X_i$, the density of $U$ can be found through convolution of the densities of $X_K$ and $\sum_{i = 1}^{K - 1}X_i$ as
\begin{eqnarray*} & {} &  f_U(x)  =  \frac{\Gamma^{K -1}(n - 1)}{\Gamma((K - 1)(n - 1))}   \\
                             & . &    \int_0^x  (n - 1)^{K - 1} y^{(K - 1)(n - 1) - 1}   (n -1) (x - y)^{n - 2}\   \,dy \\
                             & = &    (n - 1)^K  x^{K(n - 1) - 1}   \frac{\Gamma^{K -1}(n - 1)}{\Gamma((K - 1)(n - 1))} \\
                             & .  &  \int_0^1   z^{K(n - 1) - 1} (1 - z)^{n - 2}\    \,dz \\
                             & = &   (n - 1)^K  x^{K(n - 1) - 1}   \frac{\Gamma^{K}(n - 1)}{\Gamma(K (n - 1))} .
\end{eqnarray*}
The corresponding cumulative density function $F_U(x)$ is given for $x \in [0, 1]$ as
\begin{eqnarray*} F_U(x)  & = &   (n - 1)^K     \frac{\Gamma^{K}(n - 1)}{\Gamma(K (n - 1))} \frac{x^{K(n - 1) }}{K(n - 1)}  \\
& = &  \frac{\Gamma^{K}(n)}{\Gamma(K (n - 1) + 1)}    x^{K(n - 1) }   .
\end{eqnarray*}
If we define a ball in $G^K_{n,1}$ as 
\[  \tilde{B}(\delta) = \{  Q \in G^K_{n,1}\  |\   d(Q, P) \leq \delta  \}  ,\]
then its normalized volume is given by 
\begin{eqnarray*}
\mu(\tilde{B}(\delta)) &  =  & \mathrm{Pr} \{ U \leq \delta^2 \} = F_U(\delta^2) \\
                                 & = &   \frac{\Gamma^{K}(n)}{\Gamma(K (n - 1) + 1)}    \delta^{2K(n - 1) }   .
\end{eqnarray*}

\bibliographystyle{IEEEbib}
\bibliography{InterferenceAlignment-SingleCol.bbl}

\end{document}